\documentclass[a4paper,%
  12pt,%
  abstracton,%
  footexclude,%
  normalheadings,%
  pointednumbers,%
  halfparskip,%
]{scrartcl}

\newcommand{\UPLB}{University of the Philippines Los Ba\~{n}os}
\newcommand{\superscript}[1]{\ensuremath{^{\textrm{#1}}}}

\begin{document}

\title{\large Perceived Social Loafing in\\Undergraduate Software Engineering Teams}
\author{
   \small{Jaderick P. Pabico, Joseph Anthony C. Hermocilla,}\\
   \small{John Paul C. Galang and Christine C. De Sagun}\\
   \small{Institute of Computer Science}\\
   \small{\UPLB}\\
}
\date{}
\maketitle

\begin{abstract}
We surveyed 237 undergraduate students who are enrolled in various subjects and are members of software engineering teams. Their being a member in a team is part of the requirements of the course. We found that each of task visibility, distributive justice, and intrinsic task involvement were negatively associated with social loafing. We also found out that dominance, aggression and sucker effect each were positively correlated with social loafing. We further found out that perception of social loafing exists among members of software engineering teams.
\end{abstract}


\section{Introduction}
In most management of classes in the Philippines, working in teams allows students to  participate in the active construction of knowledge, enhance problem solving skills, share ideas and opinions, gain valuable experience, and learn lessons regarding group communications and problem solving that can be used later in real-world work environment. These are the reasons listed by the proponents of group work as a tool for learning in most educational systems~\cite{becker98,black02,haythornthwaite06}. In most information technology and computer science courses, this instruction technique is usually employed by instructors in courses that require the students to learn to work within a team, such as those that teach the management and control of software production teams. One of the difficulties in assessing students' output in a team is that the individual contribution is sometimes indistinguishable from the team's output~\cite{williams81}. One of the reasons for this is that without the proper tools and technology, the individual's progress is difficult to track. When a student perceives that her\footnote{Please note that we used the female gender as our writing style only, and not as a means to prejudice the opposite gender.} inputs to the team will not be given due recognition, either by her team members or by her instructor, her motivation to contribute to the team will be low~\cite{jones84}. Both the benefits of claiming high grades due to high levels of effort and the penalties of getting low grades due to low levels of effort in contributing to the team have become non-factor in motivating the students to perform better~\cite{jones84}. Because of this perception, the productivity of the team diminishes. 

One well-researched reasons for productivity losses in teams is the tendency of a team member to decrease her effort when working in a team compared to when she is working alone~\cite{williams91}. This tendency is called {\em social loafing}~\cite{latane79} and has been shown to occur in various team tasks. Work done by various researchers suggest that the occurrence of social loafing in teams are due to perceived behavioral factors of the team members. These researchers, however, have studied social loafing exclusively in laboratory settings where the environment is controlled for easy measurement of the variables. Others have conducted their research in ongoing teams, but under real-world job settings and cultures different from that of the Philippines (see for example~\cite{earley89} and~\cite{harkins80}). So far, none have studied on-going teams in the classroom settings in the Philippines. Thus, our research sought to increase understanding and awareness of social loafing as it occurs in the classroom settings and as influenced by local Filipino culture. More specifically, we wanted to determine whether the perception of  social loafing and other behavioral factors exist in the minds of students who are members of undergraduate software engineering teams. We tested hypotheses concerning the effect of the perceived behaviors of team members to the efforts exerted by a member. These behaviors were known to be antecedents to social loafing.

In this research, we surveyed 237 undergraduate students participating in at least one software engineering team from three different subjects at the \UPLB. We found out that social loafing is negatively correlated with task visibility, distributive justice, and intrinsic task involvement, and is positively correlated with dominance, aggression and sucker effect. We further found out that the perception of social loafing exists among members of software engineering teams.

The rest of the article is organized as follows. In Section~\ref{sec:framework}, we briefly discuss the origins of social loafing research and describe  behaviors that researchers consider as antecedent to social loafing. These antecedent behaviors consist the scientific basis of our investigation. In Section~\ref{sec:goals}, we state our goals for this research together with the corresponding hypotheses that we developed. In Section~\ref{sec:method}, we describe the participants, the survey and the variables that we measured in the survey. We then present the results in Section~\ref{sec:results}. Lastly, we provide a brief conclusion and discuss the  implications of our findings in Section~\ref{sec:conclude}.

\section{Scientific Framework}\label{sec:framework}
\subsection{The Origin of Social Loafing Research}

{\em Social loafing} is the term originally coined by Latane, Williams and Harkins~\cite{latane79} to refer to the tendency of a person to reduce her effort when working in teams compared to when she is working alone~\cite{williams91}. The opposite extreme of social loafing is {\em social facilitation}, a concept that people put forth more effort in the presence of others than alone~\cite{cook01}. Studies on social loafing originated from the work of Max Ringelmann in 1913, a French agricultural engineer, who was interested in the efficiency of animals, men, and machines in various agricultural applications~\cite{kravitz86}. Ringelmann observed in one of his experiments involving his students and a team of prisoners that there existed a negative correlation between the size of the team and the effort spent by each team member. This observation was later termed in his honor as the {\em Ringelmann Effect}. 

In Ringelmann's experiment involving a team of persons pulling a rope, he observed that there was a decrease in overall performance of the team when the number of members was increased. In a similar experiment involving prisoners, where each one was pulling a crank to provide manual power to a flour mill, he observed that as more men were added into the work, each man began to rely on his neighbors to perform the desired output. He further observed that some men became content to let their hand  follow the crank while some went as far as letting the crank pull their hands~\cite{kravitz86}. The former behavioral phenomenon was later termed {\em social loafer} while the latter was termed {\em free rider}.

There is a wide variety of research that exist studying the phenomenon of social loafing. Most of these research were conducted in the laboratory where the variables measured were easier to control, and variability due to errors was minimized. Most of the researchers studied on the effect of social loafing in the workplace. Research regarding social loafing in the classroom environment is relatively sparse while none exists as far as the Philippine classroom settings is concerned.

\subsection{Personal Degree of Social Loafing}

Perceived social loafing is the term used for the belief of a team member that other members are engaging in social loafing~\cite{comer95}. The central issue here is that only the perception of a team member is being measured, not the actual output of the member perceived to be engaged in social loafing. The reason for this is that there is a possibility that one member may actually struggle with the assigned concept, spend many hours of personal effort, learn a lot in the process, and yet contribute less than the others to the output of the team. Whether or not social loafing is actually occurring, our research  only measured the perception following the discourse by Mulvey and Klein~\cite{mulvey98}. This is under the assumption  that team members will base their action on the {\em perceived actions} of fellow members.

In common educational setting in the Philippines, learning actually occur within the team, but each member may perceive unequal effort. Once a member perceive that some maybe taking over the project or slacking off, it may affect her personal motivation to contribute. The act of team members carrying a free rider or social loafer has been termed playing a {\em sucker role}. Engaging in social loafing to avoid playing the sucker role is termed {\em playing the sucker effect}~\cite{kerr83}. Although a high number of research results suggest that perception of social loafing exists in teams, there has been none conducted in  the classroom settings in the Philippines. Before we conduct an investigation on the impact of social loafing to our students, we first need to know whether the perception of social loafing actually exists. 

\subsection{Individual Task Visibility}

{\em Task visibility} is defined as the belief that the class instructor is observing and tracking one's individual input to the team~\cite{kidwell93}. If the interdependence of tasks in the team is high, it is assumed that task visibility will decrease because tracking the individual contribution will be very hard~\cite{jones84}, specifically in the absence of specialized tools and appropriate technologies. When the input of a member becomes indistinguishable from the team, the member becomes unable to demonstrate her personal input and claim the benefits associated with the effort~\cite{jones84}. Moreso, a very prolific member may feel inequity and decide to social loaf if she works with other members who do not suffer the consequences of not sufficiently contributing to the team. On the other hand, she who do not fully contribute may also social loaf because she may perceive that her inputs are not critical to the team's success~\cite{karau93}. Further, she may also perceive an inequitable relationship~\cite{walster73}, believe that benefits of social loafing outweigh the cost of the penalty~\cite{murphy03}, or is intentionally {\em free riding}. Free riding is defined as the action of an individual who share the benefits of the team and yet did not spend a proportional amount of effort~\cite{albanese85}. In our study, we also wanted to know whether task visibility is a negative antecedent to the perception of social loafing in software engineering teams.

\subsection{Equity of Grade Distribution}

Liden, {\em et al.}~\cite{liden04}, defined the term {\em distributive justice} to mean as the individual's perception of the distribution of grades among team members. This is different from {\em procedural justice}~\cite{greenberg90} which is the term used for individual's perceived fairness of the procedures and policies used to compute for the grades. When participating in team activities, the achievement of a student may be influenced by her perception of procedural and distributive justice set forth by the instructor. A student might reduce her effort if there is perception of unfair distribution of the rewards~\cite{kidwell93}. Studies~\cite{liden04, karau93} show that there is a significant correlation between procedural justice and social loafing. This means that the student's perception of fairness in the procedure for grade distribution may influence the student's effort on team projects. In our study, we wanted to know whether the perceived distributive justice is an antecedent to social loafing in software engineering teams.

\subsection{Dominance and Aggression}

In any team project, the personalities of the members should be part of the design consideration of the team's composition. However, it can be expected that stronger personality types will dominate the team if no restrictions is imposed during the team's formation. The team dynamics become problematic when a member uses her position, status, or strong personality to dominate, intimidate, or harass fellow members. When members are intimidated, they usually resort to social loafing~\cite{michaelsen97}. Further, rude or angry personal attacks on a team member may result on the team member feeling unsafe, insecure and inhibited. These feelings may negatively influence the dynamics of the team~\cite{palloff03}. In our research, we wanted to know if dominance and aggression are antecedents to social loafing.

\subsection{Individual Contribution}

Researchers argue that a member of a team will likely exert extraordinary effort if she perceives that her individual effort within the team has some meaning~\cite{karau93}. When an instructor, or even a team leader, divides the work into parts, a student who was assigned the easy task may feel that she is being prejudiced and believe that her full effort is not required for the team's task. Usually, a person will withhold effort, seek to achieve personal rewards, and find ways to maximize her benefits as long as she perceives that doing so will not affect her grades~\cite{liden04}. Prejudicing the student will negatively affect her desire to perform her best. Similarly, if the member's inputs are highly integrated into the team's output while the corresponding grades are distributed accordingly, the motivation may also be affected negatively~\cite{lawler71}. Dominant members can manipulate the perception of a student's unique contribution to the team, intimidate her into believing that her efforts are not necessary, and negatively influence her desire to contribute to the team. In this study, we would like to see if dominance is an antecedent to social loafing in software engineering teams.

\section{Objectives}\label{sec:goals}

The general objectives of our study is to determine the perception of social loafing among undergraduate students who were members of a software engineering team. Specifically, we wanted:
\begin{enumerate}
\item To correlate the perceived individual task visibility and perceived social loafing;
\item To correlate the perceived individual contribution and the perceived social loafing;
\item To correlate the perceived distributive justice and the perceived social loafing; and
\item To correlate the perceived dominance and the perceived social loafing.
\end{enumerate}

\section{Methodology}\label{sec:method}

\subsection{Survey Participants}

The survey participants were 237 undergraduate students enrolled in three different courses at the \UPLB during the Summer of Academic Year (AY) 2008-2009 and First semesters of AY 2008-2009. Of the 237 students, 132 are males, 102 are females, and 3 did not report their gender. The distribution of the participants per course enrolled is as follows: 100 students were enrolled in CMSC 127 (File Processing and Database Systems~\cite{cmsc127}) during the First Semester of AY 2008-2009, 96 students were enrolled in CMSC 128 (Introduction to Software Engineering~\cite{cmsc128}) during the First Semester of AY 2008-2009, and 41 students were enrolled in CMSC 198 (Undergraduate Practicum~\cite{cmsc198}) during the Summer Semester of AY 2007-2008. These subjects required students to form a team and solve several software engineering problems throughout the semester. CMSC~127 required each team to design and implement a database system. CMSC~128  required each team to design, implement and evaluate a computerized solution to a manual processing system (e.g., inventory system, accounting system, etc.). CMSC 198 required the students to work in a host company and implement algorithmic solutions to real-world problems. Participant ages ranged between 17~to 23~years: 17 years = 1~participant; 18 years = 50; 19 years = 89; 20 years = 47; 21 years = 30; 22 years = 7; 23 years = 5; 6 participants did not report their age.

\begin{table*}[t]
\caption{Descriptive statistics from the collected survey responses. Agreed, Disagreed=percent of respondent who signified agreement/disagreement to the statements; Min=Minimum value; Max=Maximum value;SL=Social Loafing; Dist=Distributive.}\label{tab:1}
\centering
\begin{tabular}{lccccccc}
\hline\hline
Antecedent&$N$&Min&Max&Mean&SD&Agreed&Disagreed\\
\hline
1. SL-Self         &231&25&55&42.45&5.86& 1.30&74.46\\
2. SL-Others       &122& 2&39&15.05&6.36&91.80& 1.64\\
3. Task Visibility &234& 6&28&16.99&3.57&24.36&11.54\\
4. Contributions   &234& 3&13& 6.16&1.65&83.33& 0.85\\
5. Dist Justice    &234& 3&14& 6.88&2.11&63.68& 4.70\\
6. Sucker Effect   &234& 7&17&13.18&1.80& 2.99&20.51\\
7. Dominance       &227& 3&15&10.19&2.27& 9.25&42.73\\
\hline\hline
\end{tabular}
\end{table*}

\subsection{Survey Questions}

The participants were asked to voluntarily complete a survey form to report their perceptions of the following:
\begin{enumerate}
\item Degree to which their fellow team members engaged in social loafing;
\item Personal degree of social loafing;
\item Individual task visibility;
\item Individual contribution;
\item Distributive justice;
\item Sucker effect; and
\item Team member dominance.
\end{enumerate}
All survey questions were adapted from George~\cite{george92}, Piezon and Ferree~\cite{piezon08}, and Welbourne, {\em et al.}~\cite{welbourne95}. The following were the variables measured in our survey:

\begin{enumerate}
\item {\em Measurement of Perceived Team Member Loafing} - Ten survey items were asked to assess the perceived team member loafing. The survey asked the participant to indicate her perception of how many of her team members possessed the characteristics listed in the ten items.
\item {\em Measurement of Perceived Individual Loafing} - The survey also measured the participant's personal perception of her own social loafing by asking on her agreement with ten statements. The response were scaled using a five-point Likert scale (1=strongly agree and 5=strongly disagree) and were summed to form a composite.
\item {\em Measurement of Perceived Task Visibility} - The student was asked to indicate her agreement with six statements regarding her personal perception of individual task visibility. As above, the response were scaled using a five-point Likert scale.
\item {\em Measurement of Perceived Contribution} - The participant was asked to indicate her agreement with 3 statements regarding her perception of individual's contributions to the team. The response used were a five-point Likert scale as above.
\item {\em Measurement of Perceived Distributive Justice} - The participant was asked to indicate her agreement with 3 statements regarding her perception of distributive justice. The response used were a five-point Likert scale as above.
\item {\em Measurement of Sucker Effect} - Using a five-point Likert scale as above, the participant was asked to indicate her agreement with 4 statements regarding her participation in sucker effect. This measures the student's decrease in effort in response to her perceived decreased efforts by co-members.
\item {\em Measurement of Perceived Dominance} - The student was asked to indicate her agreement with 3 statements regarding her perception of a team member's dominant behavior. The response used were a five-point Likert scale as above.
\end{enumerate}

\section{Results and Discussion}\label{sec:results}

The descriptive statistics on the collected survey responses are summarized in Table~\ref{tab:1}. Of the 122 students who completed at least one item in the survey relating to social loafing in groups, 91.80\% (112 students) reported that they perceived their group members were engaging in social loafing. Of the 231 students who completed at least one item in the survey relating to social loafing by self, only 1.30\% (3 students) admitted engaging in social loafing. The main goal of our study is to determine whether the perception of social loafing exists in teams of undergraduate software engineering courses. Our results are evidence that social loafing exists. Although there is a low percentage of self-reported social loafing, it is consistent with the research results of others~\cite{karau93}. Actually, other researchers explained that the students may be unaware that they were social loafing or were just reluctant to admit that they themselves were engaging in social loafing. This observation is specially true to students from a university of known academic activism, such as the \UPLB, where students pride themselves of individual accomplishments.

Table~\ref{tab:2} shows the correlations and other corresponding statistics. From the table, it can be seen that the correlation coefficients of 10 pairs of antecedent variables are significant from zero at either $\alpha=1\%$ or $\alpha=5\%$.

\begin{table}[htb]
\caption{Correlation coefficients between all pairs of measured antecedents. Numbers in row and column headers pertain to the antecedent number in Table~\ref{tab:1};\superscript{*}coefficient is significantly different from zero at $\alpha=5\%$ ($p<0.05$); \superscript{**}coefficient is significantly different from zero at $\alpha=1\%$ ($p<0.01$).}\label{tab:2}
\centering\begin{tabular}{lcccccc}
\hline\hline
 &1&2&3&4&5&6\\
\hline
1  & \\
2  &-0.14 \\
3  &-0.25\superscript{*} &-0.06 \\
4  &-0.30\superscript{**} &-0.03    & 0.13 \\
5  &-0.15    &-0.27\superscript{**} & 0.44\superscript{**} & 0.12 \\
6  & 0.42\superscript{**} &-0.14    &-0.14    &-0.25\superscript{*} &-0.19  \\
7  & 0.43\superscript{**} &-0.08    &-0.22    &-0.27\superscript{**} &-0.17\superscript{**} & 0.38\superscript{**}\\
\hline\hline
\end{tabular}
\end{table}

\subsection{Perceived Task Visibility and Social Loafing}

Evidence from Table~\ref{tab:2} shows that the perception of decreased individual task visibility increases the social loafing by one's self ($r=-0.25$, $p<0.05$). There was no significant correlation between the perceived task visibility and the perceived social loafing by others. Our results agree with the findings of Liden, {\em et al.}~\cite{liden04} who suggested that non-recognition of an individual's input often leads to social loafing.

\subsection{Perceived Individual Contribution and Social Loafing}

The increased perception of individual contributions decreases the perceived social loafing by one's self ($r=-0.30$, $p<0.01$). The correlation between individual contributions and the social loafing by others is not significant. Again, our results support prior research~\cite{liden04}. A positive perception that one's output will be recognized will decrease the occurrence of social loafing.

\subsection{Perceived Distributive Justice and Social Loafing}

The increased perception of distributive justice decreases the perceived social loafing by others ($r=-0.27$, $p<0.01$). There was no significant correlation between the perceived distributive justice and the social loafing by one's self. Similar to Liden, {\em et al.}'s results~\cite{liden04}, our results suggest that the positive perception of the distribution of grades among members will decrease the occurrence of social loafing. This means that ensuring that the team members understand the procedures behind the grade distribution can have a positive influence on their behavior in the group. If a team member either misunderstand or perceive inequitable grade distribution, she may engage in social loafing in order to balance the perceived reward-per-effort ratio.

\subsection{Perceived Dominance vs. Social Loafing}

The perception of increased dominance increases the social loafing of one's self ($r=0.43$, $p<0.01$). There was no significant correlation between the perceived dominance and the perceived social loafing by others. The results of our study suggest that dominance negatively affects the member participation in team activities. Our result here is also supported by the positive correlation between dominance and sucker effect ($r=0.38$, $p<0.01$). As explained earlier, sucker effect is a member's reduction in effort in order to avoid carrying the load of non- or below-par-performing fellow members.

\section{Summary, Conclusion and Implications}\label{sec:conclude}

We studied the existence of social loafing among members of undergraduate software engineering teams who were enrolled in three separate courses in two semester at the \UPLB. We asked the students to voluntarily answer a survey question that will determine the occurrence of several antecedent behaviors to social loafing. We found out that social loafing exists among the team members. Based on the correlation analysis that we conducted, the following relationships were established:
\begin{enumerate}
\item There is a negative correlation between the perceived task visibility and the perceived social loafing ($r=-0.25$, $p<0.05$).
\item There is a negative correlation between the perceived individual contribution and the perceived social loafing ($r=-0.30$, $p<0.01$).
\item There is a negative correlation between the perceived distributive justice and the perceived social loafing ($r=-0.27$, $p<0.01$).
\item There is a positive correlation between the perceived dominance and the perceived social loafing $r=0.43$, $p<0.01$). There is also a positive correlation between the perceived dominance and sucker effect ($r=0.38$, $p<0.01$).
\end{enumerate}

The results of our study provide evidence that social loafing exists in undergraduate software engineering teams at \UPLB. This means that aside from the usual problems that an instructor faces in teaching software engineering-related courses, the presence of social loafing also adds to the impediment of teaching effectiveness. Thus, it is imperative that instructors and course designers consider the implications associated with social loafing when designing team projects. However, under some circumstances, social loafing may not be at all bad~\cite{jackson85}. For complex tasks, social loafing may result in reduced stress, and therefore improved performance. Instead of dealing with social loafing as a negative aspect in a team's dynamics, instructors and designers should focus on the antecedents that may lead to improved team performance. Moreover, instructors and designers should consider using currently available technologies to improve the tracking of a student's inputs. A good example is the use of the Concurrent Version System to track changes to programming codes.

\section{Acknowledgments}

We thank the Institute of Computer Science for funding this research.
\bibliography{social-loafing}
\bibliographystyle{abbrv}
\end{document}